\documentclass[11 pt, a4paper]{article}
\usepackage{jheppub}
\usepackage{amsmath}

\newcommand{\Tr}{{\rm Tr\;}}

\usepackage{graphicx}
\usepackage{amsmath}
\usepackage{amsfonts}
\usepackage{epsfig}
\usepackage{slashed}
\usepackage{braket}
\usepackage{multirow}
\usepackage{comment}
\usepackage{cases}

\def\nn{\nonumber}
\def\bec{\begin{center}}
\def\eec{\end{center}}
\def\beq{\begin{equation}}
\def\eeq{\end{equation}}
\def\bea{\begin{eqnarray}}
\def\eea{\end{eqnarray}}

\def\ahat{\hat{a}}
\def\bhat{\hat{b}}
\def\chat{\hat{c}}
\def\dhat{\hat{d}}
\def\ehat{\hat{e}}

\newcommand{\cA}{{\cal A}}

\newcommand{\cF}{{\cal F}}
\newcommand{\cFb}{{\overline{\cal F}}}
\newcommand{\cD}{{\cal D}}
\newcommand{\cDb}{{\overline{\cal D}}}
\newcommand{\cQ}{{\cal Q}}
\newcommand{\cU}{{\cal U}}
\newcommand{\cS}{{\cal S}}
\newcommand{\cN}{{\cal N}}
\newcommand{\cUb}{{\overline{\cal U}}} 

\title{Holography from lattice ${\cal N}=4$  super Yang-Mills }
\author[a]{Simon Catterall,}
\author[b]{Joel Giedt,}
\author[a] {Goksu Can Toga}

\affiliation[a]{Department of Physics, Syracuse University, Syracuse, NY 13244, USA }
\affiliation[b]{Department of Physics and Astronomy, RPI, Troy, NY 12180, USA}

\abstract{In this paper we use lattice simulation to study 
four dimensional ${\cal N}=4$ super Yang-Mills (SYM) theory. We have focused
on the three color theory on lattices of size $12^4$ and for 't Hooft couplings up to
$\lambda=40.0$. Our lattice action is based on
a discretization of the Marcus or GL twist of ${\cal N}=4$ SYM and retains one exact
supersymmetry for non-zero lattice spacing. We show that lattice theory exists in a single
non-Abelian Coulomb phase for all 't Hooft couplings. Furthermore
the static potential we obtain from correlators of Polyakov lines
is in good agreement with that obtained from holography - specifically the potential
has a Coulombic form with a coefficent that varies as the square root of the 't Hooft coupling.
}
\preprint{preprint}

\begin{document}
\maketitle

\section{Introduction}

${\cal N}=4$ super Yang-Mills is both a fascinating and non-trivial quantum field
theory. It possesses a line of conformal fixed points, is conjectured to be
invariant under a strong-weak coupling duality, and most importantly, furnishes the original example of holographic duality by providing
a description of type IIb string theory on five dimensional anti-de Sitter space.
The holographic description is most easily understood in the planar $N_c\to\infty$ strong coupling limit
where the five dimensional theory reduces to classical supergravity. However, string
loop corrections arise at $O(\frac{1}{N_c})$  and are difficult to access analytically. This provides motivation to study the theory using numerical simulation.

Naive approaches to constructing a lattice theory break supersymmetry completely
and lead to a large number of relevant supersymmetry breaking counterterms whose
couplings would need to be tuned to take a continuum limit. This stymied progress
for many years until models were constructed that preserved one or more supercharges
at non-zero lattice spacing - see the review \cite{Catterall:2009it} and references
therein. The key idea underlying these constructions is to find linear
combinations of the continuum supercharges that are nilpotent and hence compatible
with finite lattice translations. This may be accomplished either by
discretization of a topologically twisted version of the supersymmetric theory
\cite{Catterall:2007kn}
or by building a lattice theory from a matrix model using orbifolding and deconstruction techniques \cite{Cohen:2003xe,Cohen:2003qw,Kaplan:2005ta}.

While these developments have
led to many numerical studies providing evidence supporting
holography for dimensional reductions of ${\cal N}=4$ Yang-Mills 
\cite{Hanada:2016zxj,Anagnostopoulos:2007fw,Hanada:2008gy,Catterall:2008yz,Catterall:2009xn,Catterall:2010fx,Berkowitz:2016jlq,Catterall:2017lub,Rinaldi:2017mjl,Catterall:2020nmn} 
it has proven difficult until recently to test holography
directly in four dimensions.  The original
supersymmetric construction in four dimensions produced
lattice artifacts in the form of $U(1)$ monopoles that condense and 
lead to a chirally broken phase for 't Hooft couplings $\lambda >4$ \cite{Catterall:2012yq,Catterall:2014vka}. Recently we have 
constructed a new lattice action that appears to
avoid these problems 
\cite{Catterall:2020lsi}. It differs from both the original and improved \cite{Catterall:2015ira} 
actions for ${\cal N}=4$ SYM by the addition of a new supersymmetric term
that breaks the gauge symmetry from $U(N)$ to $SU(N)$. This removes the monopoles completely
and, as we will show in this paper, yields a single non-Abelian Coulomb phase.

\section{Lattice action}
We use the supersymmetric lattice action appearing in \cite{Catterall:2020lsi}.
\beq
S=\frac{N}{4\lambda} \cQ \sum_{x}\Tr \left[\chi_{ab}\cF_{ab}+\eta \left(\cDb_a\cU_a+\kappa\left({\rm Re\,det\,} \cU_a(x)-1\right)I_N\right)+\frac{1}{2}\eta d\right]+S_{\rm closed}\eeq
where the lattice field strength
\beq\cF_{ab}(x)=\cU_a(x)\cU_b(x+\ahat)-\cU_b(x)\cU_a(x+\bhat)\eeq where $\cU_a(x)$
denotes the complexified gauge field living on the lattice link running from $x\to x+\ahat$ and $\ahat$ denotes one
of the five basis vectors of an underlying $A_4^*$ lattice. $I_N$ denotes the $N\times N$ unit matrix in color space.
Similarly
\beq\cDb_a \cU_a=\cU_a(x)\cUb_a(x)-\cUb_a(x-\ahat)\cU_a(x-\ahat).\eeq
The five fermion fields $\psi_a$, being superpartners of the (complex) gauge fields, live on the 
corresponding links, while
the ten fermion fields $\chi_{ab}(x)$ are associated with new face links running
from $x+\ahat+\bhat\to x$. The scalar fermion $\eta(x)$ lives on the lattice site $x$ and is
associated with a conserved supercharge $\cQ$
which acts on the fields in the following way
\begin{align}
\cQ\, \cU_a&\to \psi_a\nn\\
\cQ\, \psi_a&\to0\nn\\
\cQ\, \eta&\to d\nn\\
\cQ\, d&\to 0\nn\\
\cQ\, \chi_{ab}&\to \cFb_{ab}\nn\\
\cQ\, \cUb_a&\to 0
\end{align}
Notice that $\cQ^2=0$ which guarantees the supersymmetric invariance of the
first part of the lattice action. The auxiliary site field $d(x)$ is needed for nilpotency of $\cQ$ offshell.
The second term $S_{\rm closed}$ is given by
\beq
S_{\rm closed}=-\frac{N}{16\lambda}\sum_x \Tr \epsilon_{abcde}\chi_{ab}\cDb_c\chi_{de}\eeq
where the covariant difference operator acting on the fermion field $\chi_{de}$ takes the form
\beq
\cDb_c\chi_{de}(x)=\cUb_c(x-\chat)\chi_{de}(x+\ahat+\bhat)-\chi_{de}(x-\dhat-\ehat)\cUb_c(x+\ahat+\bhat)\eeq

The latter term can be shown to be supersymmetric via an exact lattice
Bianchi identity $\epsilon_{abcde}\cDb_c \cFb_{de}=0$. 
Carrying out the $\cQ$ variation and integrating out the auxiliary field $d$ we obtain the final
supersymmetric lattice action $S=S_b+S_f$ where
\beq
S_b=\frac{N}{4\lambda} \sum_x\Tr\left(\cF_{ab}\cFb_{ab}\right)+
\frac{1}{2}\Tr (\cDb_a\cU_a+\kappa ({\rm Re\,det\,}\cU_a-1)I_N)^2\eeq
and
\begin{align}
S_f&=-\frac{N}{4\lambda}\sum_x \left(\Tr\chi_{ab}\cD_{\left[a\right.}\psi_{\left. b\right]}+
\Tr\eta \cDb_a\psi_a + \frac{1}{4} \Tr \epsilon_{abcde}\chi_{ab}\cDb_c\chi_{de}\right)-\\
&\frac{\kappa N}{8\lambda}\sum_{x,a}\Tr(\eta)\,{\rm det\,} \cU_a(x)\Tr(\cU_a^{-1}(x)\psi_a(x))
\end{align}
Setting $\kappa=0$ and taking the naive continuum limit $\cU_a=I+\cA_a+\ldots$  one can show
that this action can be obtained by discretization of
the Marcus or GL twist of $\cN=4$ Yang-Mills in flat space \cite{Marcus:1995mq,Kapustin:2006pk}. In the continuum this twisted formulation is used to construct a topological field
theory  but here we 
use the twisted construction merely as a change of variables that
allows for discretization while preserving the single exact supersymmetry $\cQ$. 

As described above, the discrete
theory is defined on a somewhat exotic lattice - $A_4^*$. This admits a larger
set of discrete rotational symmetries corresponding to the group
$S^5$ in comparison to those of a hypercubic lattice and this fact plays 
a role in controlling the renormalization
of the theory \cite{Catterall:2013roa}. In fact one can show that gauge symmetry, $\cQ$ supersymmetry
and $S^5$ invariance ensure that the only {\it relevant} counterterms correspond
to operators present in the classical lattice action together with a single new marginal
operator of the form $\alpha\sum_{x,a}{\rm Tr}\,(\eta \cU_a\cUb_a)$. However,
the
calculation in \cite{Catterall:2011pd} shows that even this term is
absent at $\kappa=0$ to all orders in perturbation theory.  

Finally, to retain exact supersymmetry, all fields reside in
the algebra of the gauge group -- taking their values in the adjoint representation of $U(N)$:
$f(x)=\sum_{A=1}^{N^2} T^A f^A(x)$ with $\Tr (T^A T^B)=-\delta^{AB}$.
Ordinarily this would be incompatible with
lattice gauge invariance because the measure would not be gauge invariant for
link based fields. However, in this $\cN=4$ construction the problem is evaded since the fields are 
complexified which ensures that the Jacobians that arise after gauge transformation of $\cU$ and $\cUb$ cancel.

The term involving the coupling $\kappa$ suppresses the troublesome $U(1)$ modes
while leaving the important $SU(N)$ gauge symmetry intact. It also has
another advantage; by selecting out gauge fields with unit determinant it ensures that the gauge field has the expansion $\cU_a(x)=I+\cA_a(x)+\ldots$ where $\cA_a$ are traceless complex fields.
This ensures the correct naive continuum limit.

The resultant action still possesses a set of flat directions corresponding to constant
gauge fields
that are valued in the Cartan subalgebra. To regulate these flat directions we additionally
add to the action
a term of the form
\beq
S_{\rm mass}=\mu^2\sum_x \Tr\left(\cUb_a(x)\cU_a(x)-I\right)^2\eeq
While this breaks the exact supersymmetry softly all counter terms induced by this
breaking will have couplings that are multiplicative in $\mu^2$  and hence vanishing
as $\mu^2\to 0$.

\section{Results}

Our simulations utilize the rational hybrid Monte Carlo (HMC) algorithm where the
Pfaffian resulting from the fermion integration is replaced by
\beq{\rm Pf}\,(M)={\rm det}\left[\left(M^\dagger M\right)^{\frac{1}{4}}\right]\eeq
where $M$ is the fermion
operator and the fractional power is replaced by a rational fraction approximation \cite{Clark:2006wq}. In principle this throws away any phase in the fermion
operator. In the appendix (fig.~\ref{pfaff_Nc3}) we show that this phase is always small even at
strong coupling for sufficiently small $\mu^2$.
Typical ensembles used in our analysis 
consist of $3500-4000$ HMC trajectories with $750-800$ discarded for
thermalization. Errors are assessed using a 
jackknife procedure using $20-40$ bins.
We also fix $\kappa=1$ for all
our simulations.

\begin{figure}[htbp]
\centering
\includegraphics[width=0.85\textwidth]{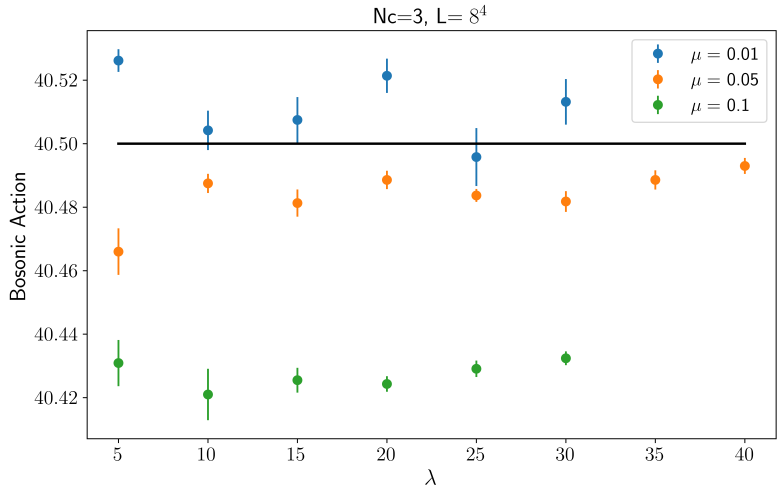} 
\caption{\label{bact_l8} Bosonic Action  vs $\lambda$ for $8^4$ lattice at $\kappa=1.0$ for various values of $\mu$}
\end{figure}

One of the simplest observables that can be measured
is the expectation value of the 
bosonic action. 
This can be calculated exactly at any coupling by exploiting
the exact lattice supersymmetry. We start by rescaling the fermions to remove the
dependence of the $\cQ$-closed term on $\beta=\frac{N}{4\lambda}$ so that the partition function $Z$ is given by
\begin{equation}
    Z=\beta^{\frac{-6N^2V}{2}}\int D\eta D\psi D\chi DU D\cUb Dd\; e^{-\beta \cQ\Lambda-\chi\cDb\chi }
\end{equation}
where $\Lambda$ generates the $\cQ$-exact terms in the action and $V$ is the number of
lattice points. After integrating
over the auxiliary field $d$ one finds that the coefficient $-\frac{6}{2}N^2V$
is shifted to $-\frac{7}{2}N^2V$. But
\begin{equation}
    -\frac{\partial\ln{Z}}{\partial\beta}=<S_B>-<S_F>=-\frac{7N^2V}{2\beta}+<\cQ\Lambda>
\end{equation}
Using the $\cQ$ Ward identity $<\cQ \Lambda>=0$ and the fact that 
the expectation value of the fermion action can be trivially found
by a scaling argument since the fermion fields appear only quadratically one finds the final result
\begin{equation}
    \frac{1}{V}\beta S_B=\frac{9N^2}{2}
\end{equation}

Fig.\ref{bact_l8} shows the bosonic action density
as a function of $\lambda$ for several values of the susy breaking mass $\mu$. It should be clear that the measured values approach the exact result $<\beta \frac{S_B}{V}>=40.5$
independent
of $\lambda$ as $\mu\to 0$. 

\begin{figure}[htbp]
\centering
\includegraphics[width=0.8\textwidth]{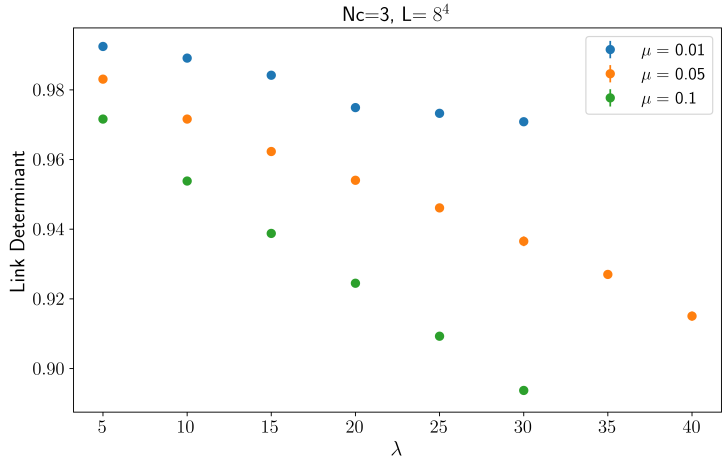} 
\caption{\label{det_l8}  Link Determinant  vs $\lambda$ for $8^4$ lattice and $\kappa=1.0$}
\end{figure}

To check that we have indeed suppressed the $U(1)$ modes we 
plot the expectation value of the link determinant in fig. \ref{det_l8}. Clearly
the observed value of the link determinant lies close to unity for
all 't Hooft couplings provided that $\mu$ is small enough. Notice that
both the bosonic action and the link determinant show no sign of a phase
transition over the range $0<\lambda<40$. This is consistent with the continuum expectation that the ${\cal N}=4$ SYM theory exists in a single phase out to
arbitrarily strong coupling. 

\begin{figure}[htbp]
\centering
\includegraphics[width=0.85\textwidth]{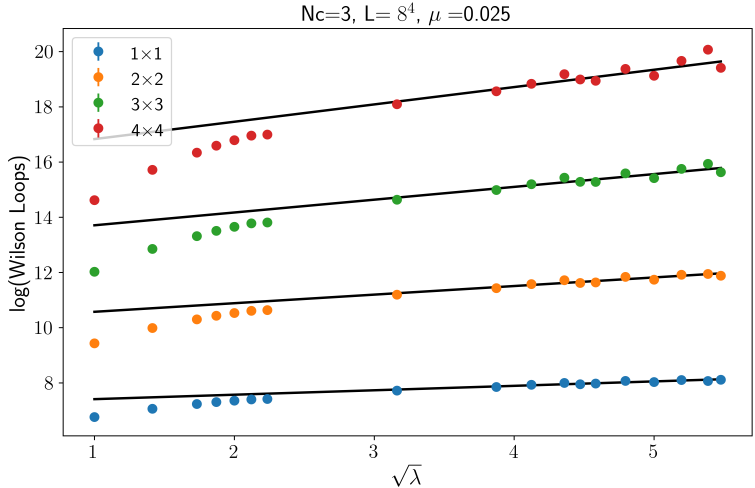} 
\caption{\label{wloop_l8}  $R \times R$ Wilson Loops  vs $\sqrt{\lambda}$ for $8^4$ lattices at $\mu=0.025$ and $\kappa=1.0$}
\end{figure}

The conclusion is strengthened further by examining a variety of supersymmetric Wilson
loop corresponding to the product of complexified lattice gauge fields
around a closed loop on the lattice. We have plotted the logarithm
of such Wilson loops
as a function of $\sqrt{\lambda}$ in fig.~\ref{wloop_l8}. Clearly the behavior
is smooth as $\lambda$ varies and again there is no sign of any phase transition.
Furthermore the linear dependence of $W$ 
seen for large $\lambda$ is consistent with holography. Indeed, both square and circular
Wilson loops can be computed in the strong coupling planar limit and show a $\sqrt{\lambda}$
dependence on the 't Hooft coupling \cite{Maldacena:1998im,Drukker:2000rr}.
It should be noted that this $\lambda$ dependence {\it cannot} be seen in perturbation theory and constitutes a non-trivial test that the lattice model
is able to reproduce the non-perturbative physics of the continuum theory.

However fig.~\ref{wloop_l8} makes it clear that there is also a perimeter
dependence to the Wilson loop. This is not unexpected and arises also in the
continuum calculations as a regulator term associated with a bare quark mass.
In general the Wilson loop arises as the amplitude for
the propagation of heavy fundamental sources that interact via
a static potential of the form
\begin{equation}
    V(r)=-\frac{\alpha}{r}+M
\end{equation}
where $M$ represents the static quark mass.
To subtract the perimeter term from our analysis and look for the presence
of an underlying non-abelian
Coulomb term in the static potential we have turned to another related
observable -- the correlation function
of two Polyakov lines. These are just Wilson lines that close via the toroidal boundary conditions. 
The measured correlator is defined by
\beq
P(r)=\sum_{x,y} \left[<P(x)P^\dagger(y)>-<P(x)><P^\dagger(y)>\delta(r,|x-y|)\right]\eeq
where $|x-y|$ is the distance in the $A_4^*$ lattice.
As for the Wilson loop it is expected to vary like
\beq
P(r)\sim e^{-V(r)t}\eeq
with $V(r)$ the static potential.
\begin{figure}[htbp]
\centering
\includegraphics[width=0.85\textwidth]{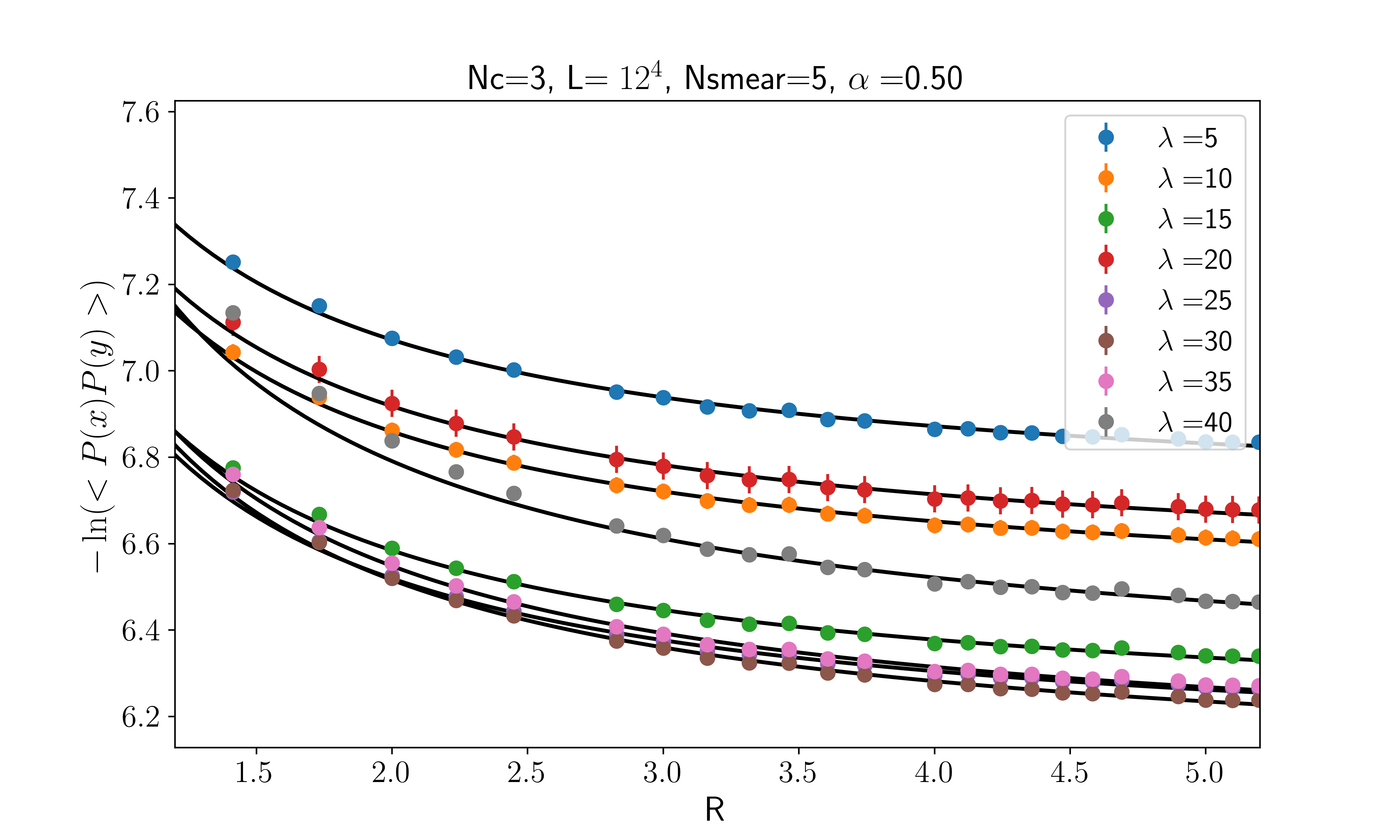} 
\caption{\label{pc_n5_a0.50}  $-\ln{<P(x)P(y)>}$ vs $R$ for $12^4$ lattices at $\mu=0.05$ and $\kappa=1.0$}
\end{figure}
In practice we have computed this correlator on ensembles of
{\it smeared} Polyakov lines. Smearing the lattice gauge fields
is a procedure for replacing each of the lattice gauge links by an average
over neighboring link paths or staples. This smearing
procedure has the effect of reducing U.V effects
and increasing the signal to noise ratio in observables that depend on
the gauge field. It also suppresses the contribution of any bare quark mass.
We have used the APE smearing procedure which replaces the link fields $\cU$ in the following way \beq
\cU_a(x)^{(N_{smear})} = (1 -\alpha)\cU_a(x)^{(N_{smear}-1)} +\frac{\alpha}{8(1-\alpha)} \cS_a(x)^{(N_{smear}-1)}
\eeq
 where $\cS$ denotes the sum over directional staples, $N_{smear}$ denotes the number of iterative smearing steps and $\alpha$ is the smearing coefficient \cite{albanese1987glueball}.
 
We have fitted the correlator for a range of smearing parameters at each value of
$\lambda$ assuming a non-abelian Coulomb form 
for $V(r)$ -- see the
tables \ref{fit_l12_n5_a0.45}-\ref{fit_l12_n5_a0.55} in the appendix. In practice we
find that $N_{smear}=5$ and $0.45 \leq \alpha \leq 0.55$ yields good, robust fits to the data
over the range $2<r<5$ on a $12^4$ lattice.
Fig. \ref{pc_n5_a0.50} shows the logarithm of this correlator for a lattice of size $L=12^4$ with $\mu=0.05$ with $N_{smear}=5, \alpha=0.50$.

\begin{figure}[htbp]
\centering
\includegraphics[width=0.85\textwidth]{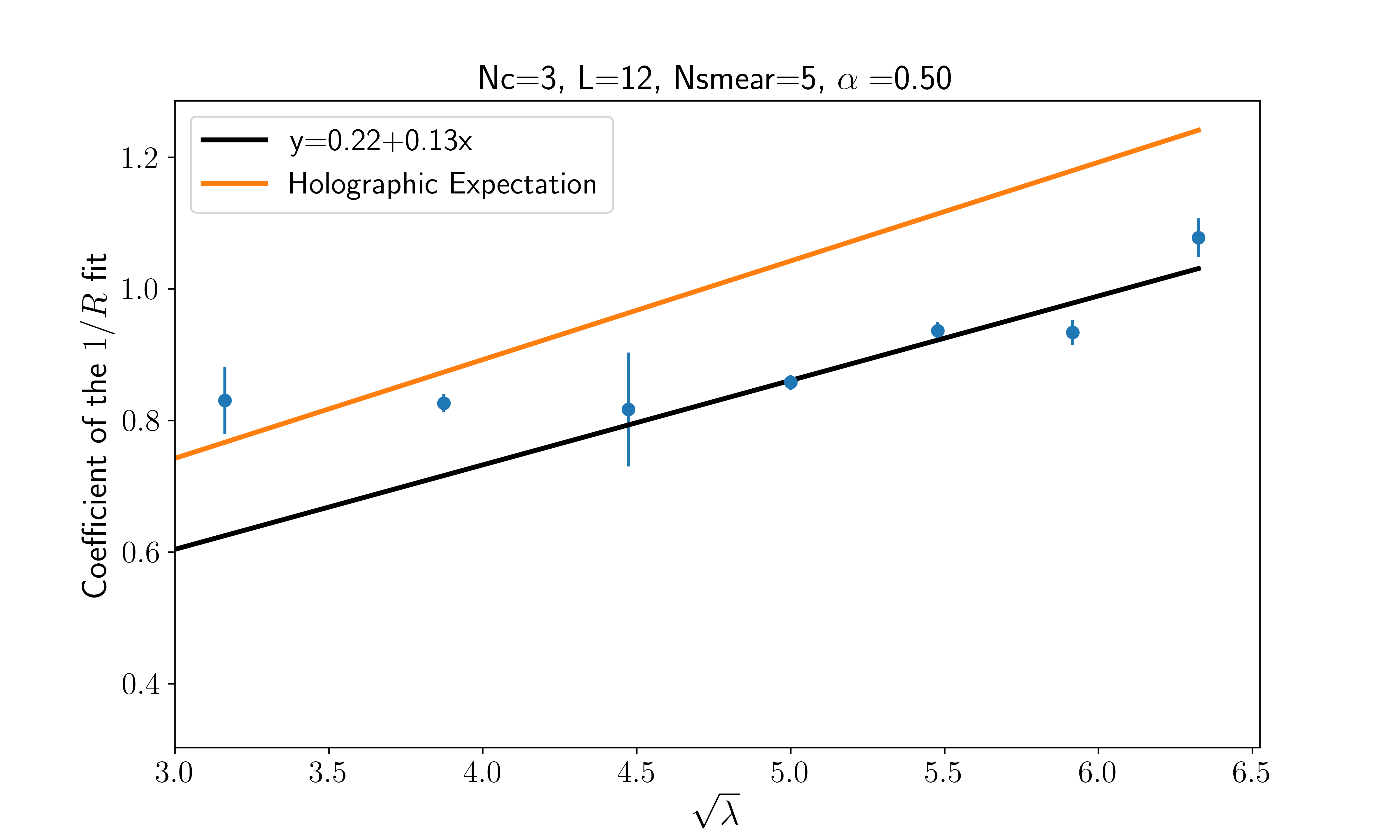} 
\caption{\label{const_L12_a0.50}  Coefficient of the $1/r$ vs $\sqrt{\lambda}$ for $12^4$ lattices at $\mu=0.05$}
\end{figure}

Taking the coefficients from these Coulomb fits and plotting them as a function of 
$\sqrt{\lambda}$ we again see a linear dependence on $\sqrt{\lambda}$
which is consistent with the holographic
expectation \cite{Chu:2009qt} . Indeed, even the numerical coefficient in the fit
lies within $10\%$ or so of the holographic prediction which can be seen in fig.\ref{const_L12_a0.50}. Note that the holographic prediction has been expressed in terms of the lattice coupling $\lambda$ and not the continuum coupling.  
In the appendix \ref{ap_fit} we include equivalent fits (figs.\ref{const_L12_a0.45} and \ref{const_L12_a0.55}) for a range of different smearing parameters thereby verifying that the agreement with the holographic prediction is robust.

\section{Conclusions}
We have studied a new supersymmetric lattice action for ${\rm N}=4$
super Yang-Mills in four dimensions at strong 't Hooft coupling. 
We have focused on the case of
three colors $N=3$ and utilized lattices as large as $12^4$. 
Correlators of (smeared) Polyakov lines show that the static potential exhibits
a non-Abelian Coulomb form $V(r)=\frac{\alpha \sqrt{\lambda}}{r}$ where
the value of $\alpha$ and the square root dependence
on the 't Hooft coupling match expectations 
from holography.

\acknowledgments
This work was supported by the US Department of Energy (DOE), Office of Science, Office of High Energy Physics, 
under Award Numbers {DE-SC0009998} (SC,GT) and {DE-SC0013496} (JG). Numerical calculations were carried out on the DOE-funded USQCD facilities at Fermilab and NSF-funded ACCESS SDSC-Expanse facilities under Award number PHY170035.

\bibliographystyle{JHEP3}
\bibliography{susy}

\appendix

\newpage
\section{Appendix - Phase of the Pfaffian}

\begin{figure}[!htbp]
\centering
\includegraphics[width=0.8\textwidth]{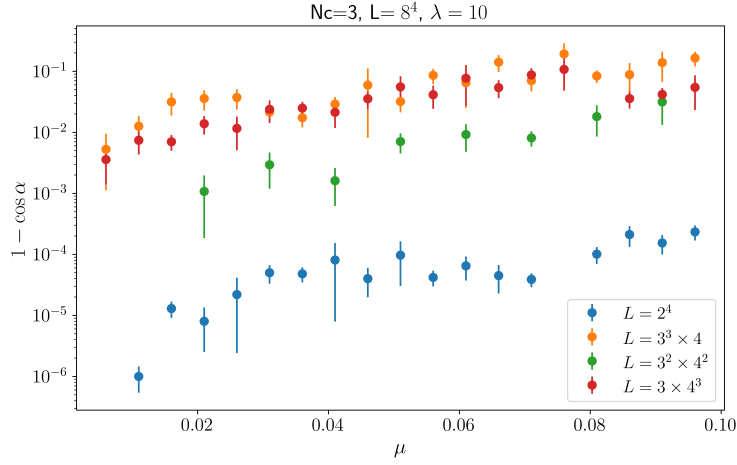} 
\caption{\label{pfaff_Nc3} $1-cos(\alpha)$ vs $\mu$ for $L=2^4,3^3\times 4, 3^2\times 4^2, 4^3\times 3$}
\end{figure}

\section{Appendix -- Dependence of the fits on smearing parameters}\label{ap_fit}
\begin{figure}[!htbp]
\begin{minipage}[b]{0.55\linewidth}
\centering
\includegraphics[width=\textwidth]{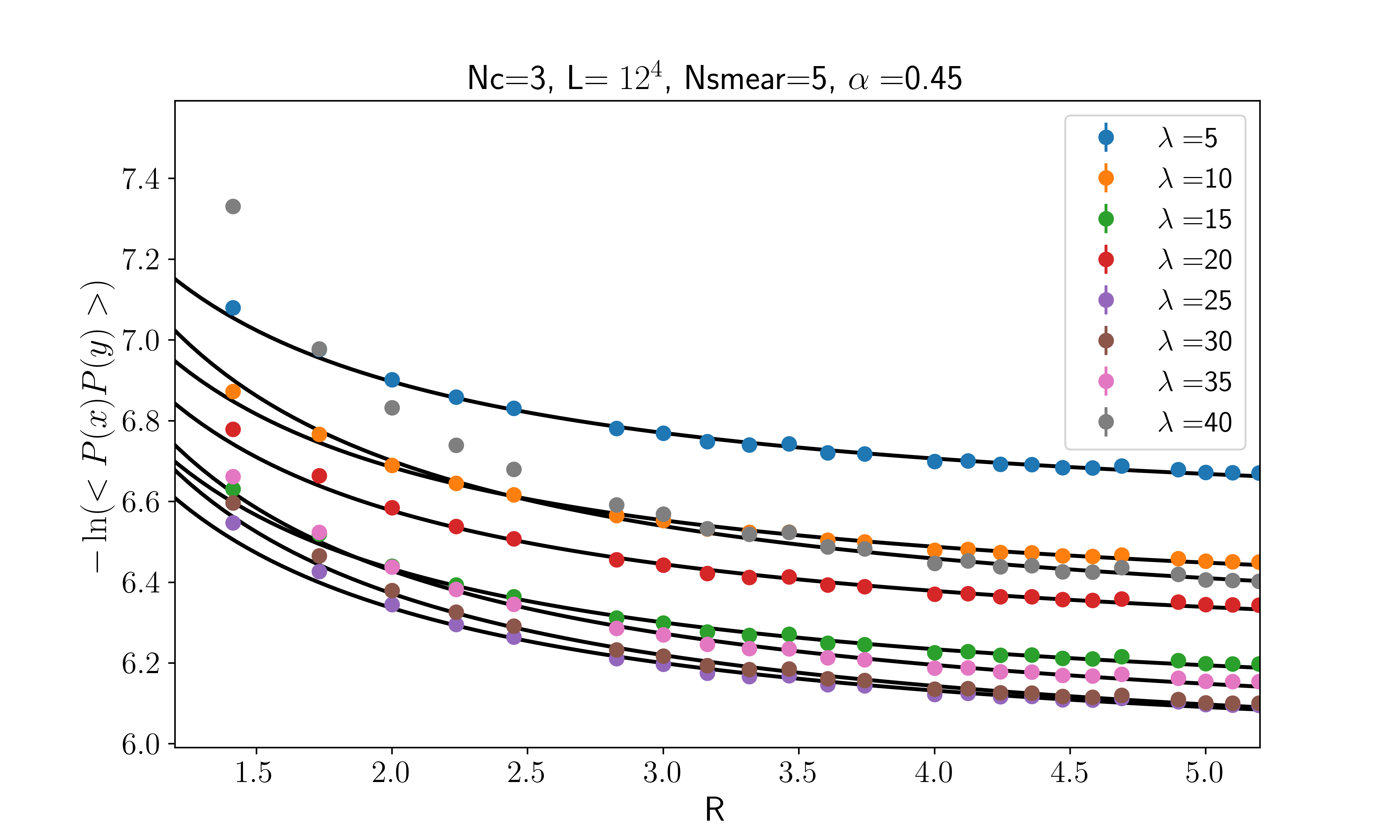}
\caption{$-\ln{<P(x)P(y)>}$ for $\alpha=0.45$}
\label{pc_n5_a0.45}
\end{minipage}
\hspace{0.5cm}
\begin{minipage}[b]{0.55\linewidth}
\centering
\includegraphics[width=\textwidth]{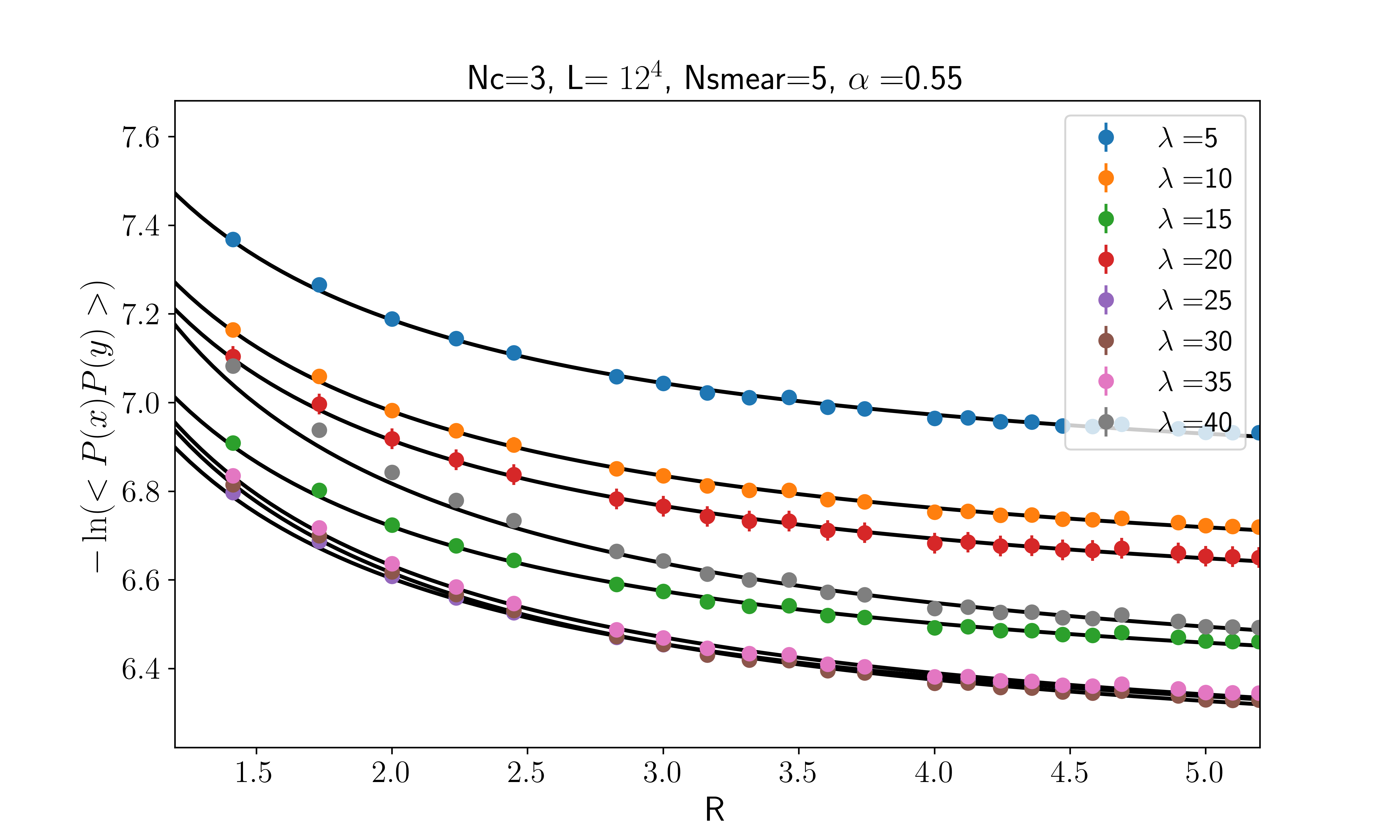}
\caption{ $-\ln{<P(x)P(y)>}$ for $\alpha=0.55$}
\label{pc_n5_a0.55}
\end{minipage}
\end{figure}

\begin{figure}[!htbp]
\begin{minipage}[b]{0.55\linewidth}
\centering
\includegraphics[width=\textwidth]{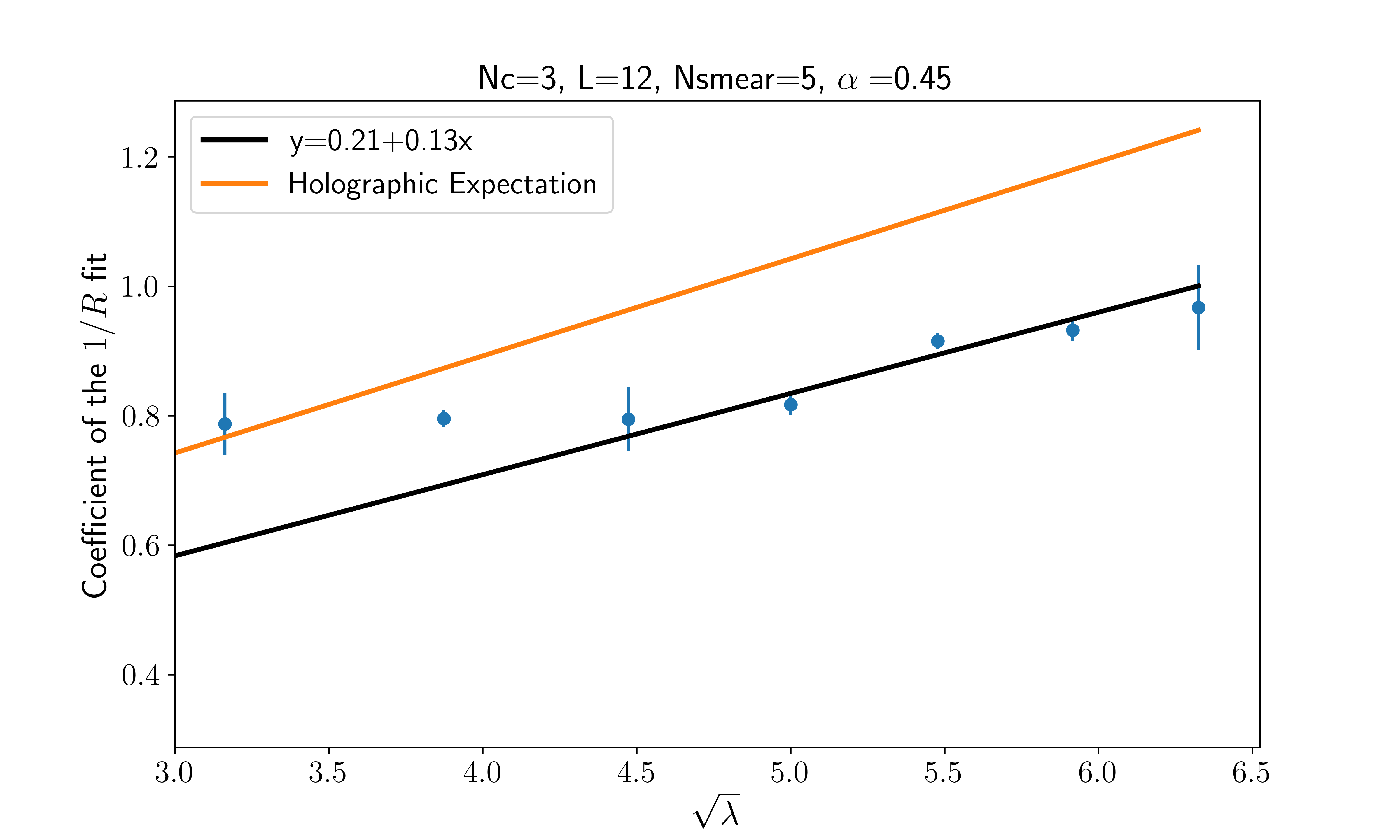}
\caption{\label{const_L12_a0.45}  Coefficient of the $1/r$ vs $\sqrt{\lambda}$ for $\alpha=0.45$}
\end{minipage}
\hspace{0.5cm}
\begin{minipage}[b]{0.55\linewidth}
\centering
\includegraphics[width=\textwidth]{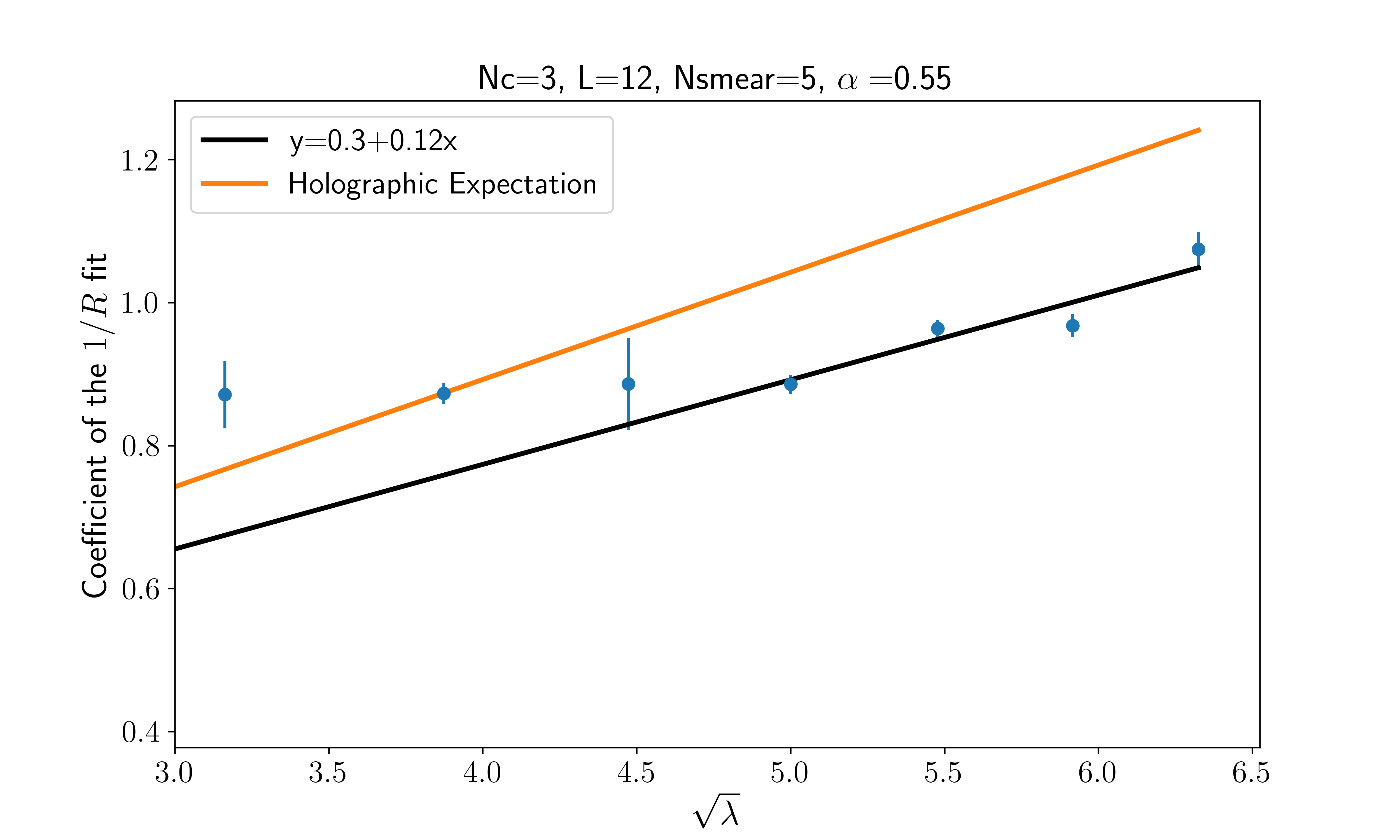}
\caption{ Coefficient of the $1/r$ vs $\sqrt{\lambda}$  for $\alpha=0.55$}
\label{const_L12_a0.55}
\end{minipage}
\end{figure}


\label{fitting_nsmear}

\begin{table}[ht]
	\centering
	\begin{tabular}{|c|c|c|}
		\hline
		$\lambda$ &$a+b/R$ & Reduced-$\chi^2$ \\ \hline
		$5$   & $6.51(1) + 0.76(3)/R$   & 0.18         \\ \hline
		$10$ & $6.29(1) +0.78(5)/R$    & 0.085           \\ \hline
		$15$ & $6.03(4) + 0.79(1)/R$    & 1.6          \\ \hline
		$20$ & $6.06(1) +0.80(5)/R$   & 0.13          \\ \hline
		$25$ & $5.92(1) + 0.81(2)/R$   & 1.4           \\ \hline
		$30$ & $5.91(1) +0.92(1)/R$   & 1.8         \\ \hline
		$35$ & $5.96(1) +0.93(2)/R$   & 1.2          \\ \hline
            $40$ & $6.22(2) +0.97(1)/R$ ($3.5<R<5.0$)  & 2.3          \\ \hline
	\end{tabular}
	\caption{\label{fit_l12_n5_a0.45} $1/R$ Fitting results for $L=12^4,\mu=0.05,N_{smear}=5,\alpha=0.45$  }
\end{table}

\begin{table}[htbp]
	\centering
	\begin{tabular}{|c|c|c|}
		\hline
		$\lambda$ &$a+b/R$ & Reduced-$\chi^2$ \\ \hline
		$5$   & $6.67(1) + 0.79(4)/R$   & 0.16         \\ \hline
		$10$ & $6.44(2) +0.83(5)/R$    & 0.07           \\ \hline
		$15$ & $6.17(1) + 0.82(1)/R$    & 1.50          \\ \hline
		$20$ & $6.51(3) +0.81(8)/R$   & 0.04           \\ \hline
            $25$ & $6.09(1) +0.86(2)/R$   & 2.10           \\ \hline
		$30$ & $6.09(1) + 0.937(1)/R$   & 1.34          \\ \hline
		$35$ & $6.05(1) +0.934(2)/R$   & 0.92         \\ \hline
		$40$ & $6.25(1) +1.07(2)/R$ ($3.0<R<5.0$)  & 3.03        \\ \hline
            
	\end{tabular}
	\caption{\label{fit_l12_n5_a0.50} $1/R$ Fitting results for $L=12^4,\mu=0.05,N_{smear}=5,\alpha=0.50$  }
\end{table}

\begin{table}[htbp]
	\centering
	\begin{tabular}{|c|c|c|}
		\hline
		$\lambda$ &$a+b/R$ & Reduced-$\chi^2$ \\ \hline
		$5$   & $6.76(1) + 0.85(3)/R$   & 0.17         \\ \hline
		$10$ & $6.54(1) +0.87(5)/R$    & 0.09           \\ \hline
		$15$ & $6.28(1) + 0.87(1)/R$    & 1.23          \\ \hline
		$20$ & $6.47(2) +0.88(6)/R$   & 0.06           \\ \hline
            $25$ & $6.09(1) +0.88(1)/R$   & 1.53          \\ \hline
		$30$ & $6.13(1) + 0.96(1)/R$   & 1.47          \\ \hline
		$35$ & $6.14(1) +0.97(2)/R$   & 0.90         \\ \hline
		$40$ & $6.28(1) +1.07(2)/R$ ($2.5<R<5.0$)  & 2.48        \\ \hline
            
	\end{tabular}
	\caption{\label{fit_l12_n5_a0.55} $1/R$ Fitting results for $L=12^4,\mu=0.05,N_{smear}=5,\alpha=0.55$  }
\end{table}

\end{document}